# THE GREEN GRID SAGA -- A GREEN INITIATIVE TO DATA CENTERS: A REVIEW

PARTHA PRATIM RAY*

*Department of Electronics and Communication Engineering,
Haldia Institute of Technology, Haldia, Purba Medinipur-721657, West Bengal, India
parthapratimray@hotmail.com*

**Abstract**

Information Technology (IT) significantly impacts the environment throughout its life cycle. Most enterprises have not paid enough attention to this until recently. IT's environmental impact can be significantly reduced by behavioral changes, as well as technology changes. Given the relative energy and materials inefficiency of most IT infrastructures today, many green IT initiatives can be easily tackled at no incremental cost. The Green Grid - a non-profit trade organization of IT professionals is such an initiative, formed to initiate the issues of power and cooling in data centers, scattered world-wide. The Green Grid seeks to define best practices for optimizing the efficient consumption of power at IT equipment and facility levels, as well as the manner in which cooling is delivered at these levels hence, providing promising attitude in bringing down the environmental hazards, as well as proceeding to the new era of green computing. In this paper we review the various analytical aspects of The Green Grid upon the data centers and found green facts.

*Keywords*: Green computing; data center; green grid.

## 1. Introduction

Environmental issues have been given the most priority in recent years to Information Technology (IT) domain especially on data centers. A data center [30] is such a facility, used to house computer systems and associated components, such as and storage systems and telecommunications. It generally includes redundant or backup power supplies, redundant data communications connections, environmental controls (e.g., air conditioning, fire suppression) and security devices.

Energy use is a central issue for data centers. Some facilities have power densities more than 100 times that of a typical office building [9]. For higher power density facilities, electricity costs are a dominant operating expense and account for over 10% of the total cost of ownership (TCO) of a data center [13]. By 2012 the cost of power for the data center is expected to exceed the cost of the original capital investment [17]. In August of 2007, the Environmental Protection Agency (EPA) published a report to congress on"Server and Data Center Energy Efficiency". This report detailed the rapidly growing energy costs of data centers ($4.5 billion in 2006, $7.4 billion projected by 2011) and the dramatic increase in data center power consumption (61 billion kWh in 2006, 100 billion projected by 2011. In that same year entire information and communication technologies (ICT) sector was estimated to be responsible for roughly 2% of global carbon emissions with data centers accounting for 14% of the ICT footprint [18]. The EPA estimates that servers and data centers are responsible for up to 1.5% of the total United States electricity consumption [19], or roughly 0.5% of US GHG emissions, for 2007 and Given a business as usual scenario green house gas emissions from data centers is projected to more than double from 2007 levels by 2020 [20].

The Green Grid (TGG) [23] is a global consortium of companies, government agencies and educational institutions dedicated to advancing energy efficiency in data centers and business computing ecosystems. The Green Grid does not endorse vendor-specific products or solutions, and instead seeks to provide industry-wide recommendations on best practices, metrics and technologies that will improve overall data center energy efficiencies.

In this review paper we first analyze the various metrics implied by TGG on data centers to make them energy efficient. Then we look for the power consumption and carbon dioxide emission attributes by data centers and the improvement of these parameters in contrary to as TGG implementations. Thirdly we concentrate on cooling management and unused server related techy implications by TGG. Fourthly on impacts of virtualization. At last we present the corporate structural changes needed to make data centers green with the help of data center logical design guide.

This paper is organized as follows: Section 2 presents different TGG metrics. Section 3 presents the energy consumption and carbon dioxide emission scenario. Section 4 presents cooling and server management issues. Section 5 represents impact of virtualization. Section 6 represents the corporate structure and green

---

* Netaji Road, By lane – 2, Newtown, Post + District- Cooch Behar, Pin- 736101, West Bengal, India.

data center design guide.

## 2. Metrics

TGG found that several metrics can help IT organizations better understand and improve the energy efficiency of their existing datacenters, as well as help them make smarter decisions on new datacenter deployments. They are as below.

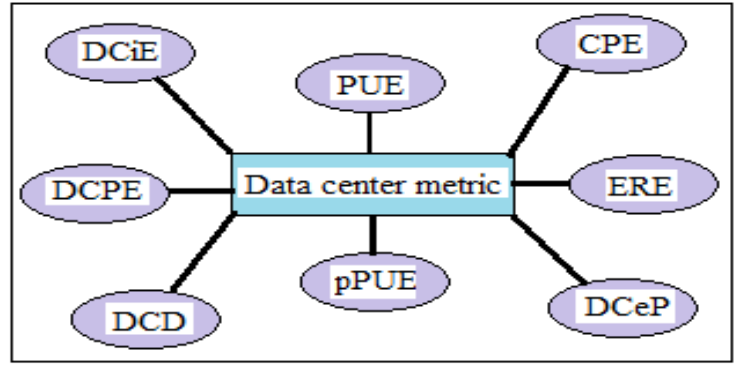

Fig. 1. Different metrics to measure data center efficiency.

### 2.1. Tactical

TGG propose the use of Power Usage Effectiveness (PUE) [14] and Datacenter Efficiency (DCE) [14] Rassmussen or Data Center infrastructure Efficiency (DCiE) metrics, which enable data center operators to quickly estimate the energy efficiency of their data centers, compare the results against other data centers, and determine if any energy efficiency improvements need to be made as in Figure 1 below.

- PUE = Total Facility Power / IT Equipment Power   (2.1)
- DCiE = IT Equipment Power / Total Facility Power   (2.2)

TGG also considers the development of metrics that provide more granularity for the PUE metric by breaking it down into the following components: [10]

### 2.2. Strategic

Another Data Center Performance Efficiency (DCPE) metric and a refined version of the PUE metric adopted for all major power-consuming subsystems in the datacenter is described as follows: [4]

- DCPE= Useful Work / Total Facility Power   (2.3)

*Total Facility Power* is measured at or near the facility utilitys meter(s) to accurately reflect the power entering the data center. *IT Equipment Power* would be measured after all power conversion, switching, and conditioning is completed and before the IT equipment itself. The *PUE* can range from 1.0 to infinity. Ideally, a *PUE* value approaching 1.0 would indicate 100% efficiency (i.e. all power used by IT equipment only). Currently, there are no comprehensive data which show the true spread of the PUE for datacenters. Some preliminary work indicates that many datacenters may have a PUE of 3.0 or greater, but with proper design a PUE value of 1.6 should be achievable [4].

### 2.3. DCD

While there are metrics today used to gauge the performance of the data center, their usefulness falls short when measuring data center performance-per-watt. One such metric that is Data Center Density (DCD): [24]

- DCD = Power of All Equipment on Raised Floor / Area of Raised Floor  (2.4)

### 2.4. CPE

TGG has introduced an interim metric or proxy for productivity to allow data centers today to estimate their productivity as a function of power used named Compute Power Efficiency [10].

- CPE = (IT Equipment Utilization  IT Equipment Power) / Total Facility Power  (2.5)
- CPE = IT Equipment Utilization / PUE  (2.6)

### 2.5. *ERE*

Energy Reuse Effectiveness (ERE), which will provide the data center practitioner with greater visibility into energy efficiency in data centers that make beneficial use of any recovered energy from the data centers [16].

- ERE = (Cooling + Power + Lighting + IT-Reuse) / IT  (2.7)
- ERE = c × PUE  (2.8) where c is a factor determined below.

Energy Reuse Factor:
- ERF = Reuse Energy / Total Energy  (2.9) ERE can then be defined as:
- ERE = (1-ERF) × PUE  (2.10)

It can be seen that as ERF goes to 0 (no energy reuse), ERE will equal PUE, as one would expect for a well-behaved metric. It could represent a very efficient design (PUE = 1.2) with a small amount of energy reuse (ERF = 0.17 and ERE = (1-0.17) × 1.2 = 1.0) or an inefficient base design (PUE = 2.0) with a lot or energy reuse (ERF = 0.50 and ERE = (1-0.5) × 2.0 = 1.0). The theoretical limits of ERE, and ERF can be summarized as: 0<= ERE<= ∞ and 0<= ERF<= 1.0

### 2.6. *DCeP*

TGG introduced the concept of a family of DCP [11] metrics (referred to as DCxP metrics) based on a definition that measured useful work produced in a data center divided by any quantization of a resource consumed to reduce that work. The first metric in this family, DCeP, where the resource consumed is energy. Thus, DCeP is defined as follows:

- DCeP = Useful work produced in a data center / Total energy consumed in the data center to produce that work (2.11)

### 2.7. *pPUE*

The Partial PUE (pPUE) is a new conceptual metric where a PUE-like value for a subsystem (i.e. container is an eligible subsystem) can be measured and reported.

### 3. Energy consumption and $CO^2$ emission

Since the early2000 a large number of data centers have been implemented in different countries to fill up human needs through computational work, firing up the twins (i.e. power consumption and CO2 emission) again in picture. Figure 2-left, charts the scenarios in the power consumption projection from 2007 to 2011 showing TGG the best of all. In the February 2009 survey, the participating Tier I (less number of servers) data centers were the least likely to measure PUE, and those that did reported the highest average PUE values (Figure 2-right) [2].

We know that server power consumption scales linearly with CPU utilization. Hence altering server processor power saving features such as performance state or p-state (The combination of a specific CPU frequency and voltage) a server's power consumption can be reduced (Figure 3-left). Figure 3-right, shows the chart of power saving and $CO_2$ (Million Metric Ton) reduction projection in 2011.

## 4. Cooling and server management

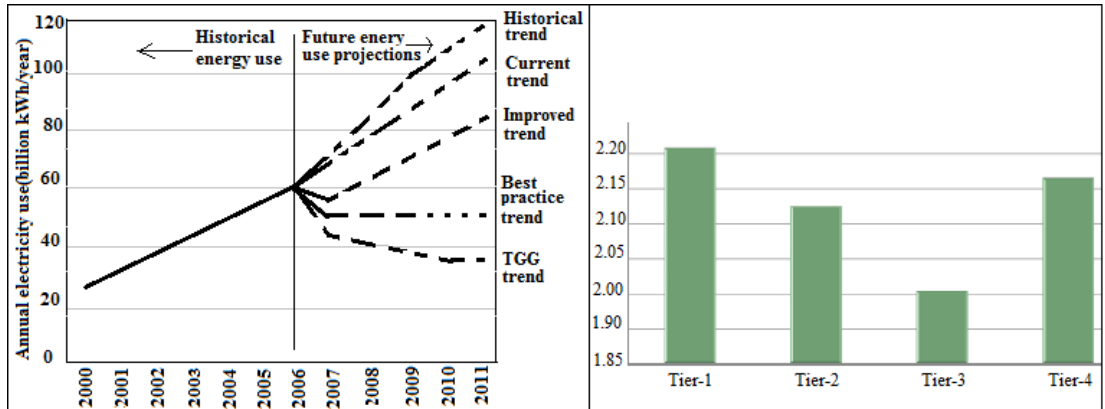

Fig. 2. Comparison of projected electricity use, all scenarios, 2007 to 2011 [27] (left), Tier I data centers showed the highest PUE measurements (right).

In 2007, the Technology and Strategy Work Group of TGG consortium established a system architecture Task Force to investigate and provide guidance on existing and emergent cooling technologies along with the deep understanding of data center layout and operation that can improve the efficiency of data center cooling architectures. Computational Fluid Dynamics (CFD) analysis can be used to optimize data center airflow by identifying weak areas of cooling capacity in the data center. Figure 4-left shows an example of a CFD analysis. The cut-plane across the space shows the air temperature in the space. Figure 4-right represents different cooling solutions [12].

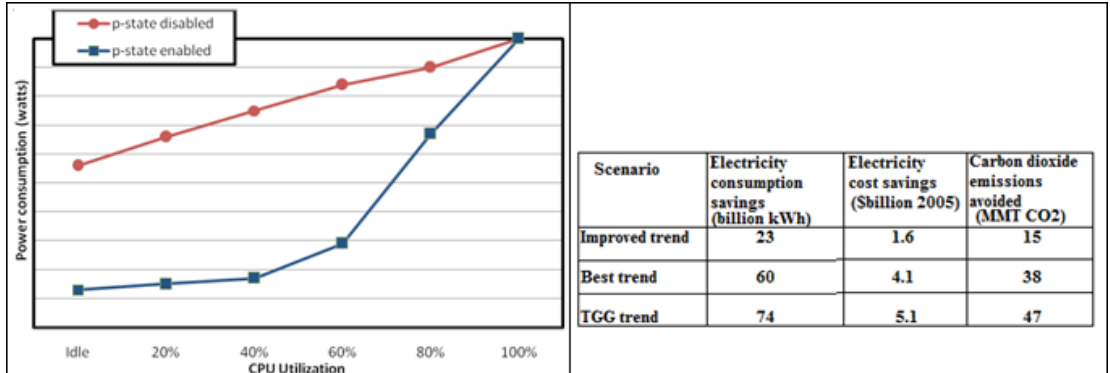

Fig. 3. Impact of p-state on power consumption [5] (left), Annual Savings in 2011 by Scenario (Compared to Current Efficiency Trends) [26] (right).

In December 2009, TGG conducted a brief, seven-question online survey to learn more about the number and nature of data center servers that are left on but perform no useful work. The organization heard from 188 respondents, primarily data center managers in the United States, who represented a range of industries and server counts. The survey results above tells that almost 80% of the respondents reported that they have unused servers in their data centers. Figure 5 shows the number of unused server in various IT organizations.

According to a 2009 report [15] from analyst firm International Data Corporation (IDC), organizations worldwide in 2008 attributed approximately $145,000,000,000 to management and administration of 33 million servers. That works out to be more than $4,000 per server estimated on 4.4 million unused servers.

A March 2010 online article [29] notes that licensing fees and storage expenses for unused virtual servers can tie up resources that could be deployed elsewhere. It cites one organization that discovered it was paying approximately $50,000 in disk and licensing costs for 42 virtual servers that had been offline for more than 90

days. That example indicates that an unused virtual server can lead to roughly $1,000 in wasted disk and license costs.

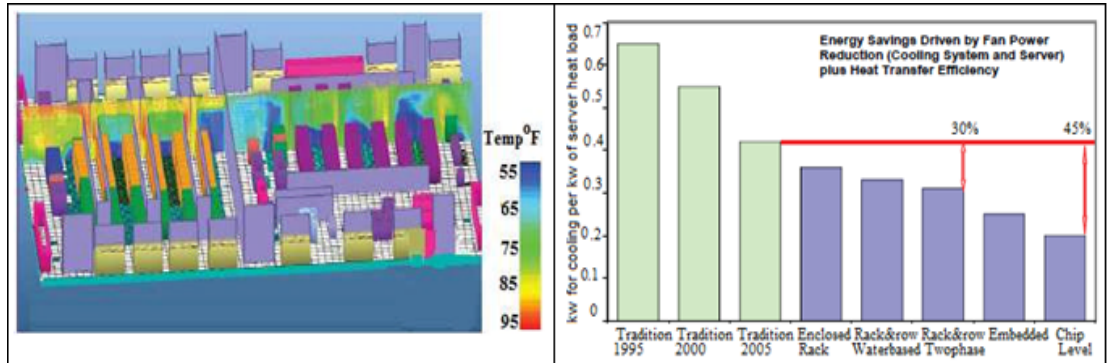

Fig. 4. Example of an enhanced multi-airflow design[12] (left), Relative efficiency potentials of different cooling solutions (right).

## 5. Impact of virtualization

A primary motivation for consolidation using virtualization is increased computing efficiency more computing per watt of power consumed by the data centers. The reduced power consumed by the consolidated IT load itself, however, is only the beginning of the savings entitlement that can be claimed when consolidating workloads to a smaller set of physical equipment. The two pie charts illustrates that the fan power represents a larger proportion of the electrical bill after virtualization. Therefore, even though the electric bill decreased (smaller pie chart) it did not decrease (Figure 6-left) by the same amount as the server power. Since virtualization and the resulting equipment consolidation, can significantly reduce load, over sizing is an important efficiency issue in a virtualized data center (Figure 6-right).

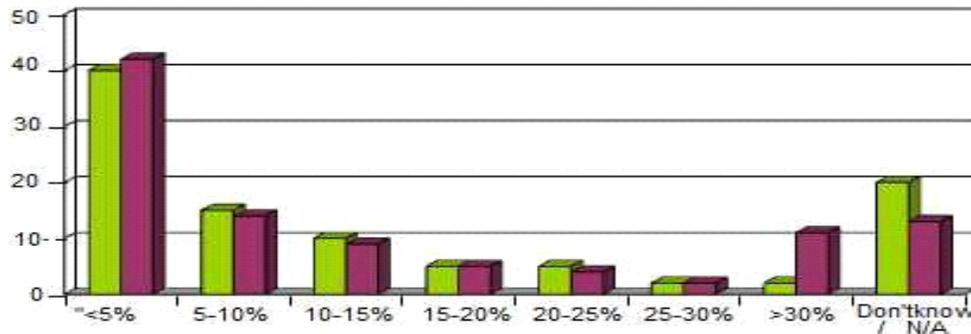

Fig. 5. Comparison of the number of unused physical versus physical and virtual servers found in respondents' organizations [6].

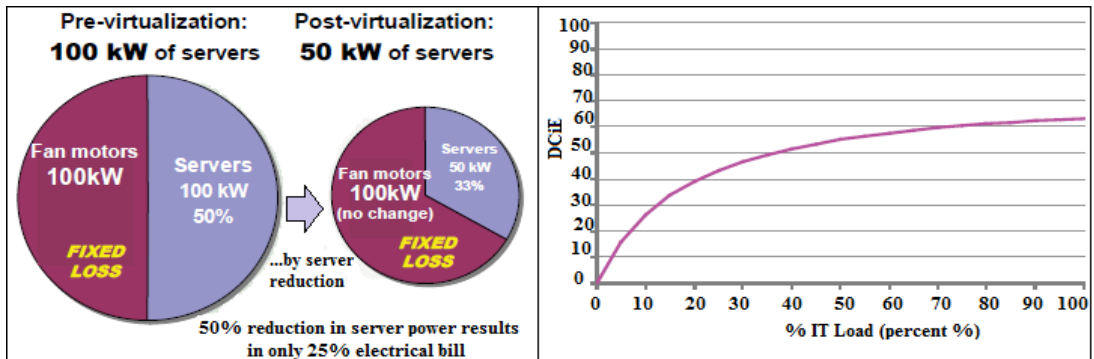

Fig. 6. Electricity Savings vs. Efficiency Gains (Fans Servers only) (left) and Typical infrastructure efficiency curve, using DCiE as the metric (right)[8].

## 6. Corporate structure and data center design guide

While new technologies enable IT organizations to significantly increase the delivered computational performance per watt, the effects of expanding demand for growth is being felt in other parts of the organization, especially Facilities departments. This section explains how greater cooperation by these departments and an integrated, holistic, collaborative approach to their respective missions can greatly enhance how data centers are designed, built, and operated so as to maximize resources and provide the levels of service needed by IT. These strategic and tactical IT and Facilities issues are summarized in Figure 7.

|  | IT | Facilities |
|---|---|---|
| STRATEGIC | CIO<br>• IT as an innovator for the Line of Businesses<br>• Business Continuity Plan<br>• Computing, storage, and network strategies<br>• Data Center Consolidation<br>• Regulatory issues | SVP Corporate Real Estate<br>• Data Center Consolidation<br>• Leasing or purchasing<br>• Outsourcing or Insourcing<br>• Regulatory issues<br>• Corporate sustainability initiatives<br>• Availability of electric utilities<br>• Energy costs |
| TACTICAL | Data Center Manager<br>• Implementing technology<br>• IT Service Level Agreements<br>• IT operational efficiencies | Facility Manager<br>• Providing adequate and reliable power and cooling<br>• Facilities Service Level Agreements<br>• Facility operational efficiencies |

Fig. 7. Strategic and tactical issues for IT and facilities managers.

Typical organizational structures have IT and Facilities in different parts of the organization with little or no common metrics. To improve this, Facilities must be viewed by senior management as a close partner of IT and more integral to the enterprises overall success. Management must change its view of Facilities as in Figure 8. In 2009, TGG has proposed the logical structure of data center design as a design guide towards the data centers as shown in the Figure 9.

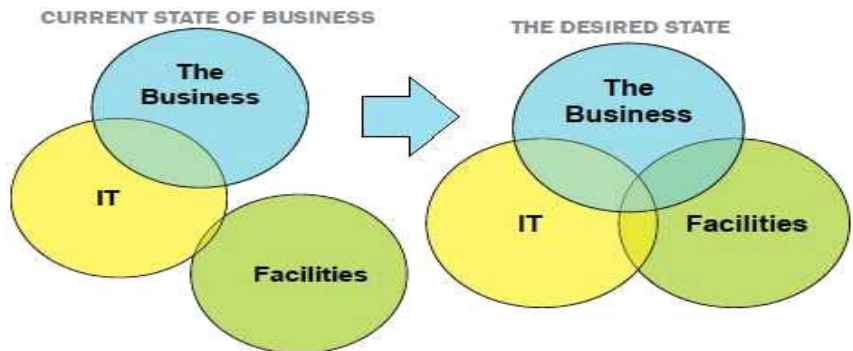

Fig. 8. Different states of business [25].

## 7. Conclusion

In this paper we present the review on The Green Grid a non-profit trade organization of IT professionals formed to address the issues of power and cooling in datacenters. We have collected and analyzed the information about The Green Grid and its approaches based upon power, CO2, cooling, virtualization, corporate and data center design guide. Currently, we are investigating the various other parameters to make the data centers green, hence a more step towards the green computing.

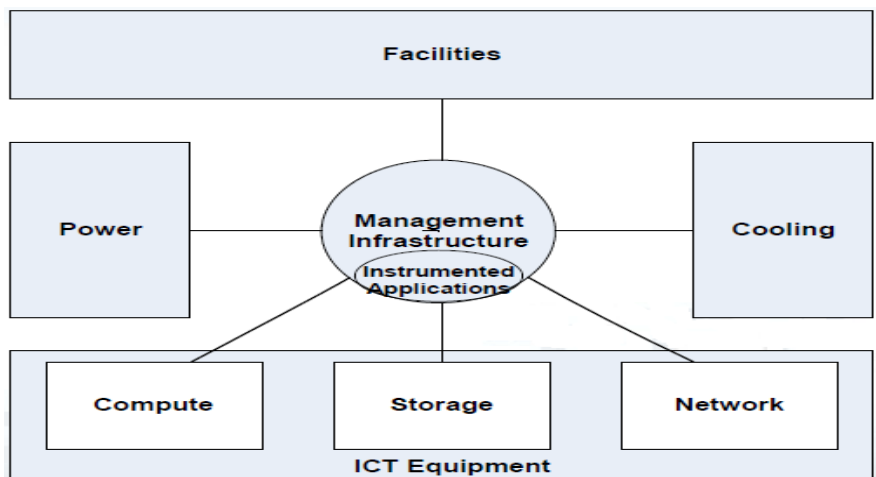
Fig. 9. Different states of business [25].